\documentclass{svproc}
\usepackage{url}
\usepackage{color}
\usepackage{graphicx}
\RequirePackage{xspace}
\newcommand{\gev}{\ensuremath{\mathrm{\,Ge\kern -0.1em V}}\xspace}
\newcommand{\gevc}{\ensuremath{{\mathrm{\,Ge\kern -0.1em V\!/}c}}\xspace}

\begin{document}
\mainmatter              
\title{Particle Identification with the TOP and ARICH detectors at Belle II}
\titlerunning{Particle Identification at Belle II}  
%
\author{S Sandilya \\ (On behalf of the Belle II Collaboration)}
\authorrunning{S Sandilya} 
\institute{University of Cincinnati, Cincinnati, Ohio 45221.\\
  \email{saurabhsandilya@gmail.com}.
}

\maketitle              

\begin{abstract}
  Particle identification at the Belle II experiment will be provided by two ring imaging Cherenkov devices, the time of propagation counters in the central region and the proximity focusing RICH with aerogel radiator in the forward end-cap region.
  The key features of these two detectors, the performance studies, and the construction progress is presented.
  \keywords{Particle identification, Cherenkov radiation, Ring Imaging Cherenkov Counter}
\end{abstract}
\section{Introduction}
A reliable particle identification (PID) is important for any high energy physics experiment, and in case of $B$-factories it is inevitable~\cite{krizan:2009}.
In the $B$-factories, PID is required to tag $B$-meson flavour for CP violation studies in the neutral $B$-meson system, and to suppress backgrounds in precision measurements of rare $B$ and $D$ decays.
In Belle II detector~\cite{belle2:2011}, PID will be performed by the Time-Of-Propagation (TOP) counter in the central region and the Aerogel Ring Imaging Cherenkov (ARICH) counter in the forward endcap region.
The working principle of both the TOP and ARICH counters is based on imaging the Cherenkov rings. 
In these proceedings, we report the design, method and present status of the PID systems in both the regions of the Belle II detector.

\section{TOP counters}
\label{top}
A TOP counter module primarily consists of a quartz radiator bar, micro-channel plate photomultipliers (MCPPMTs) and a front-end readout. 
Two quartz bars each having dimension $(1250 \times 450 \times 20)~\rm mm^{3}$, a mirror with dimension $(100 \times 450 \times 20)~\rm mm^{3}$, and a small expansion prism of dimension $(100 \times 456 \times 20-51)~\rm mm^{3}$ are epoxied to form the radiator.
The radiator is enclosed in a box made of aluminum honeycomb panels and is supported by PEEK polymer buttons.
The mirror focuses parallel rays in the Cherenkov cone to a single point and thus removes the effect of the bar thickness and also allows to correct for chromatic dispersion.
The quality of the optical components of the radiator is ensured by several quality acceptance tests~\cite{boqun}.

Cherenkov photons emitted by a charged track in the radiator go through total internal reflections and registered at the expansion volume end by an array of MCPPMTs.
The arrival time, including the time of flight of the charged particle and the time of propagation of emitted photons, and position in the detection plane of each Cherenkov photon are used to compute a likelihood for a given particle mass hypothesis~\cite{ratcliff}.

A square-shaped MCPPMT has been developed in collaboration with Hamamatsu Photonics KK for the TOP counter~\cite{inami}\cite{akatsu}.
It has a multi-alkali photocathode whose average quantum efficiency is about 28\% for wavelengths around 380 nm. By applying the nominal operating bias, a gain of $10^{6}$ is achieved.
Each module contains two rows of 16 MCPPMTs, giving in total 32 MCPPMTs per module. 
The MCPPMTs are physically mounted on the front-end electronics modules. 
The main components of the front-end electronics are: front board to host the MCPPMT array; high voltage board to provide high voltages to the MCPPMTs; Application-Specific Integrated Circuits (ASICs); and Standard Control, Read-Out, and Data (SCROD) board~\cite{andrew:ieee}\cite{andrew:tipp2014}.

A small prototype of the TOP counter was tested at the $1.2~\rm GeV/c$ positron beam at LEPS (Laser Electron Photon beam line at SPring-8) in June 2013 and results were found in agreement with the MC expectations. 

\section{ARICH counters}
\label{epid}

The ARICH counter is basically a proximity focusing RICH. Its main components are: aerogel tiles as a radiator, an array of position sensitive photon detectors, and a readout system~\cite{matsumoto}\cite{iijima}\cite{nishida}. 
The aerogel radiator consists of two layers with increasing refractive indices along the particle path so that the Cherenkov photons emitted by each layer overlap at the photo-detection plane. 
The two aerogel tiles are of thickness 20~mm each and have refractive indices of 1.045 and 1.055 for upstream and downstream tiles, respectively. 
This arrangement of two layers with different refractive indices gives better performance than a single aerogel layer for the entire thickness~\cite{korpar}.
The photo-detector for the ARICH should be sensitive to single photon detection, able to provide position information, be immune to the 1.5~T magnetic field perpendicular to the photon detection plane, and be tolerant of the high radiation environment. 
A Hybrid Avalance Photo-Detector (HAPD) was developed in a joint effort with Hamamatsu which has a peak quantum efficiency of about 28\% at 400~nm.
The bombardment gain is of about 1800 and an additional gain of about 40 is achieved from the avalanche process of the APDs, resulting an overall gain of 70,000.
The ARICH covers $3.5~\rm m^{2}$ in the forward endcap of the Belle II detector; the aerogel radiator plane and the photo-detection plane are separated by a distance of 200~mm.
The radiator plane consists of 124 pairs of aerogel tiles while the photo-detection plane contains 420 HAPDs. At the outermost edge of these two planes, planar mirrors are placed to redirect the outside-going photons towards the detection plane.
Dedicated high gain and low noise electronics were developed for the readout.
To each HAPD a front-end board with four ASICs and a field programmable gate array is attached.
The digitized hit information is collected by a merger board from front-end boards, and then communicated to further stages of the data acquisition system~\cite{nishida:readout}.

A prototype ARICH was tested at the DESY test beam in May 2013. The Cherenkov angle resolution is found to be 15.8~mrad, and on average 9 photons per track are detected. 
\section{Summary and Status}
\label{summary}
The TOP and ARICH detectors will provide PID information in the barrel and forward endcap region, respectively, of the Belle II detector. 
The working principle of both the detectors is based on imaging the Cherenkov ring created by the passage of charged particles. 
The TOP utilizes the impact position and time of arrival of the Cherenkov photons at the detection plane after total internal reflections in the quartz radiator bar to obtain PID information, whereas the ARICH is a proximity focusing RICH detector.
Prototypes of both the detectors demonstrated their expected performance in test-beams. 
All 16 TOP counters have been successfully installed in the Belle II detector.
ARICH detector is under construction now and will be finished by the spring of 2017 and its installation to the Belle II detector is expected in the summer.
Detailed simulations have been performed with the Belle II software framework, based on which we expect an excellent charged kaon efficiency ($>90\%$) with a very small ($<10\%$) charged pion misidentification probability.
%
%


\begin{thebibliography}{20}
%
\bibitem{krizan:2009}
  Krizan P, 2009
  {\it Journal of Instrumentation} {\bf 4}, P11015.

\bibitem{belle2:2011}
  Abe T{\it et al.} (Belle II Collaboration), 
  arXiv:1011.0352.
  
\bibitem{boqun}
  Wang B 2013 
  {\it  Nucl. Instrum. Methods Phys. Res., Sect.}~A {\bf 766} 204--207.
  
\bibitem{ratcliff}
  Ratcliff B N 2003 
  {\it  Nucl. Instrum. Methods Phys. Res., Sect.}~A {\bf 502} 211--221.

\bibitem{inami}
  Inami K 2008 
  {\it  Nucl. Instrum. Methods Phys. Res., Sect.}~A {\bf 595} 96--99.
  
\bibitem{akatsu}
  Akatsu M 2004 
  {\it  Nucl. Instrum. Methods Phys. Res., Sect.}~A {\bf 528} 763--775.

\bibitem{andrew:ieee}
  Andrew M 2012 
  {\it  IEEE Realtime Conf. Rec.} 1--5.
  
\bibitem{andrew:tipp2014}
  Andrew M 2014 
  {\it  PoS}~{\bf (TIPP2014)} 171. 
  
\bibitem{matsumoto}
  Matsumoto T {\it et al.} 2004 
  {\it  Nucl. Instrum. Methods Phys. Res., Sect.}~A {\bf 521} 367--377.
  
\bibitem{iijima}
  Iijima T {\it et al.} 2005 
  {\it  Nucl. Instrum. Methods Phys. Res., Sect.}~A {\bf 548} 383--390.
  
\bibitem{nishida}
  Nishida S {\it et al.} 2014 
  {\it  Nucl. Instrum. Methods Phys. Res., Sect.}~A {\bf 766} 28--31. 
  
\bibitem{korpar}
  Korpar S {\it et al.} 2007 
  {\it  Nucl. Instrum. Methods Phys. Res., Sect.}~A {\bf 572} 429--431.

\bibitem{nishida:readout}
  Nishida S {\it et al.} 2012
  {\it  Phys. Procedia} {\bf 37} 1730--1735.
  
\end{thebibliography}
\end{document}